\documentclass[floatfix,superscriptaddress,twocolumn,aps,prl]{revtex4-2}
\usepackage{dcolumn}
\usepackage{graphicx}
\usepackage{bm}
\usepackage{physics}
\usepackage[mathlines]{lineno}
\usepackage{setspace}
\usepackage[utf8]{inputenc}
\usepackage[english]{babel}
\usepackage[T1]{fontenc}
\usepackage[dvipsnames]{xcolor}
\usepackage{amssymb}
\usepackage{amsmath}
\usepackage{hyperref}
\usepackage{amsmath}
\usepackage{xcolor}
\usepackage{chemist}
\usepackage{graphicx}
\graphicspath{ {./images/} }
\usepackage{chemformula}
\usepackage{orcidlink}

\begin{document}

\title{Volume Collapse Without a Structural Transition in Shock-Compressed FeO}

\author{C. Crépisson\email{celine.crepisson@physics.ox.ac.uk}\orcidlink{0009-0005-7482-252X}}\affiliation{Department of Physics, Clarendon Laboratory, University of Oxford, Parks Road, Oxford OX1 3PU, UK}

\author{T. Stevens}
\affiliation{Department of Physics, Clarendon Laboratory, University of Oxford, Parks Road, Oxford OX1 3PU, UK}

\author{M. Fitzgerald\orcidlink{0009-0004-5716-1996}}\affiliation{Department of Physics, Clarendon Laboratory, University of Oxford, Parks Road, Oxford OX1 3PU, UK}

\author{C. Camarda\orcidlink{0000-0002-1702-9817}}\affiliation{European XFEL, Holzkoppel 4, 22869 Schenefeld, Germany}\affiliation{Fakultät Physik/Dortmund Electron Accelerator, Technische Universität Dortmund, Dortmund, Germany}

\author{P. G. Heighway\orcidlink{0000-0001-6221-0650}}\affiliation{Department of Physics, Clarendon Laboratory, University of Oxford, Parks Road, Oxford OX1 3PU, UK}

\author{D. Peake}
\affiliation{Department of Physics, Clarendon Laboratory, University of Oxford, Parks Road, Oxford OX1 3PU, UK}

\author{D. McGonegle\orcidlink{0000-0001-5329-1081}}\affiliation{AWE, Aldermaston, Reading, RG7 4PR, United Kingdom}

\author{A. Descamps\orcidlink{0000-0003-1708-6376}}\affiliation{School of Mathematics and Physics, Queen’s University Belfast, University Road, Belfast BT7 1NN, UK}

\author{A. Amouretti\orcidlink{0000-0001-8114-613X}}\affiliation{The University of Osaka, Graduate School of Engineering Science, 2-1 Yamada-oka, Suita, Osaka 565-871, Japan}

\author{D. A. Chin}\affiliation{University of Rochester, Laboratory for Laser Energetics (LLE), 250 East River Road Rochester NY, 14623-1299, USA}

% ...

\date{\today}

\begin{abstract}
We report x-ray diffraction and emission spectroscopy of FeO under laser-driven shock compression between 31–199 GPa. FeO retains the B1 (rocksalt) structure along the Hugoniot to the melt boundary at 191~GPa. While the phase and volume are broadly consistent with results from static compression, we observe an anomalous 7–10\% volume collapse around 60~GPa absent in static experiments. We identify this as an isostructural high-spin to low-spin metallic transition in FeO. The low-spin state is directly evidenced by x-ray emission spectroscopy at 180~GPa.

\end{abstract}

\maketitle

Fe$_{1-x}$O (with x between 0.04 and 0.12 at ambient condition~\cite{hazen}) is a transition metal oxide of central relevance to the Earth’s lower mantle. Ferropericlase, (Fe,Mg)O, is a major constituent of this region, which spans pressures of approximately 25-136~GPa~\cite{dziewonski}. Seismological observations have identified 5-40~km-thick ultralow-velocity zones (ULVZs) at the base of the D'' region directly above the core–mantle boundary (CMB). These zones are characterized by reduced seismic velocities accompanied by increased density~\cite{yugarnero,hansen,lay1998}. 
Iron oxides, including Fe$_{1-x}$O, are therefore potential contributors to the physical properties of the CMB region. Large volumes of iron oxides may be delivered to the lowermost mantle through the subduction of Banded Iron Formations (BIFs), which formed by oceanic iron precipitation following the emergence of oxygenic photosynthesis. Subducted BIFs have been proposed as the source of some ULVZs~\cite{dobson}, as well as iron oxides formed by oxidation reactions involving water released from subducted slabs at the CMB~\cite{mao}. In addition, the sound velocity of Fe$_{1-x}$O at CMB pressures is significantly lower than that of other lower-mantle minerals, consistent with the reduced seismic velocities inferred for ULVZs~\cite{tanaka_2020}. Electronic transitions (spin and metallisation) are another important aspect of iron-bearing minerals, which affect lower mantle density, conductivity, and seismic behaviour. As a result, Fe$_{1-x}$O has been widely studied at the pressure–temperature conditions of the Earth’s lower mantle and at the higher pressures relevant to the mantles of super-Earth exoplanets~\cite{coppari}.

\begin{figure}
     \centering
     \includegraphics[width=1\columnwidth]{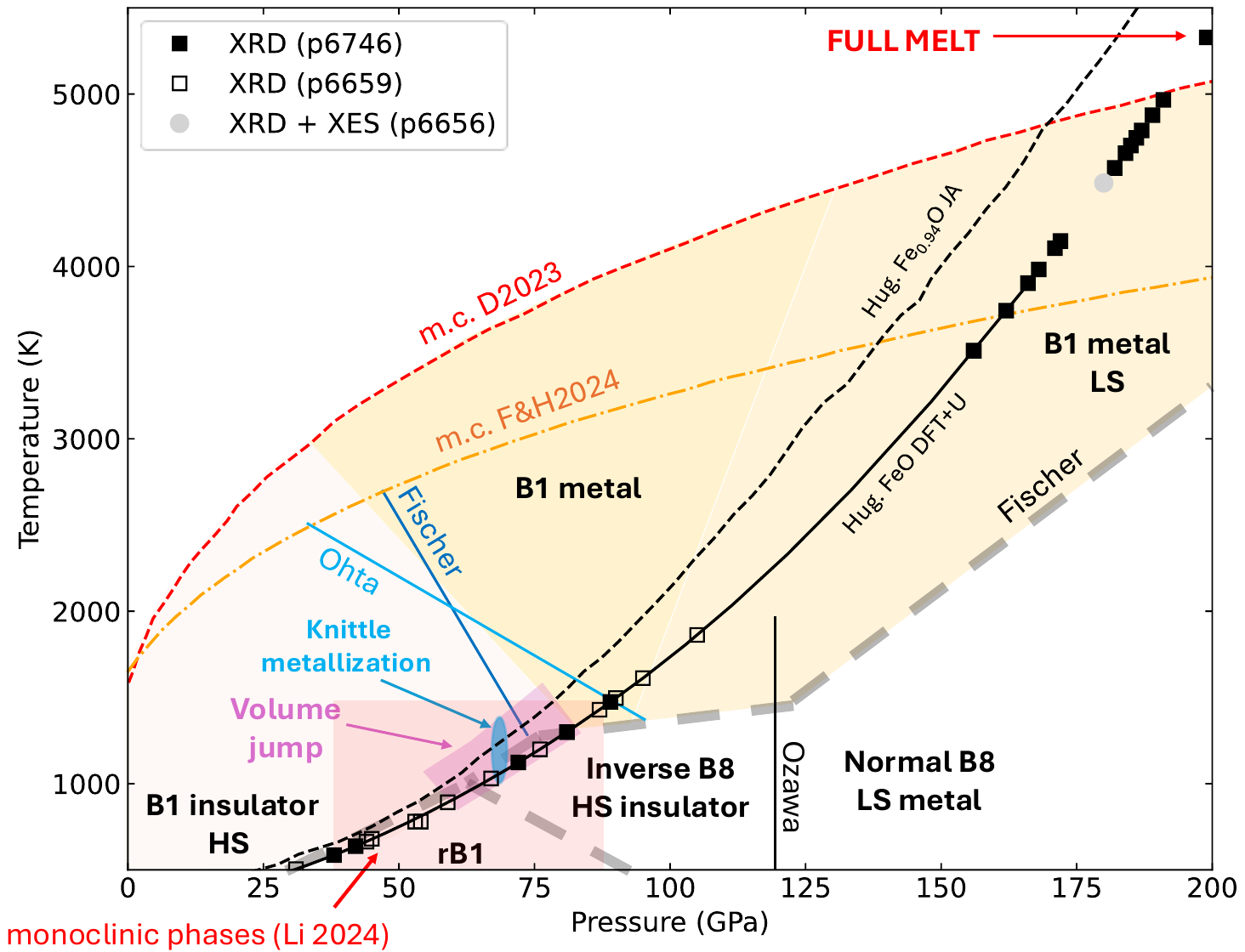}
     \caption{Phase diagram of Fe$_{1-x}$O. Phase boundaries (thick dashed grey lines) are taken primarily from Fischer~{\it et al.}~\cite{fisher}, based on experiments in the Fe–FeO system, for which FeO is expected to be close to stoichiometric above 5~GPa~\cite{stolen,seagle}. Insulator–metal transition boundaries are derived from studies of Fe$_{1-x}$O~\cite{fischer_met,ohta}, consistent with reported rB1 phase relations~\cite{fei_mao} and melting curves (m.c.) from two recent studies: D2023~\cite{dobrosavljevic} and F\&H2024~\cite{fu_hirose}. Solid black curve is our DFT+U Hugoniot (Hug.) with present experimental results. Hugoniot calculated from experimental results is also shown (JA)~\cite{jeanloz}. The B8 phase boundaries shown here differ from those of Fei and Mao~\cite{fei_mao} and lie at slightly lower temperatures than those reported by Ozawa {\it et al.}~\cite{ozawa_2010}. Monoclinic phases have been reported within the red shaded region~\cite{li_FeO,kantor}.
     } 
     \label{Fig:phase}
\end{figure}

Under high-pressure and high-temperature conditions, Fe$_{1-x}$O exhibits several crystalline phases, including the B1 rocksalt structure, a rhombohedrally distorted variant of B1 (rB1)~\cite{fei_mao,mao_1996,jacobsen}, and the B8 hexagonal NiAs-type structure~\cite{fei_mao,murakami,ozawa_2010,fisher}, as shown in Fig.~\ref{Fig:phase}. At pressures above 240 GPa, a B2 CsCl-type phase has also been reported~\cite{ozawa,coppari}. More recently, single-crystal x-ray diffraction (XRD) studies have identified monoclinic phases at pressures as low as 39~GPa and at elevated temperatures~\cite{kantor,li_FeO}.
Li {\it et al.}~\cite{li_FeO} proposed that this monoclinic phase may remain stable up to melting and to pressures of at least 136~GPa. In this interpretation, diffraction peak splitting observed prior to melting, and previously attributed to a defect-driven order–disorder transition~\cite{dobrosavljevic}, can be instead explained by monoclinic symmetry. As a consequence, some Bragg reflections previously assigned to the B8 phase could alternatively be indexed as monoclinic.
The resulting discrepancies in reported B8 phase boundaries, and in the identification of monoclinic phases, have been attributed to differences in Fe$_{1-x}$O stoichiometry~\cite{seagle} and potentially to kinetic effects~\cite{fisher}. In this work, we therefore focus on the Fe–FeO system, which is directly relevant to the CMB and allows stabilization of nearly stoichiometric FeO above 5~GPa~\cite{stolen,seagle}. In this system, static compression studies consistently report the B1, rB1, and B8 phases~\cite{fisher,seagle,murakami,ozawa_2010,campbell}, with the B1 structure stable under lower-mantle pressure conditions~\cite{fisher}.

Under lower-mantle pressure–temperature conditions, Fe$_{1-x}$O undergoes an insulator–metal transition~\cite{fischer_met,ohta,knittle_met}, as shown in Fig.~\ref{Fig:phase}. At high temperature, a quantum critical state has been proposed to persist up to 150~GPa~\cite{ho}, potentially contributing to the high electrical conductivity inferred for the CMB. Fe$_{1-x}$O also exhibits a pressure-induced spin transition from high-spin to low-spin (HS–LS) states~\cite{greenberg_FeO,ozawa_2011}. This transition has been invoked as a possible source of seismic heterogeneities within the lower mantle~\cite{lin_2005,cohen} and, potentially, near the top of the outer core~\cite{cohen,hamada}.
The pressure at which the spin transition occurs, as well as the pressure range over which it is completed, remain debated and appear to depend on crystal structure and temperature. For FeO B1, calculations predict the onset of the transition near 70~GPa at 2000~K, with completion only at pressures approaching 200~GPa~\cite{ohta_2012}. Greenberg {\it et al.} reported a similar transition pressure of 73~GPa at 1160~K~\cite{greenberg_FeO}, consistent with earlier calculations by Leonov {\it et al.}~\cite{leonov}, but also reported x-ray emission spectroscopy (XES) signatures characteristic of the high-spin state at 90~GPa and 2000~K. This contrasts the results of Ohta {\it et al.}~\cite{ohta_2012}, suggesting that the transition pressure may increase at high temperature.

XES and M\"ossbauer measurements, conducted primarily within the stability fields of the rB1, B8, or monoclinic phases, place the onset of the spin transition between 85 and 90~GPa~\cite{pasternak_FeO,hamada,li}. For the B8 phase at 1500~K, a higher transition pressure of 120~GPa has been reported~\cite{ozawa_2011}. In contrast, Badro {\it et al.}~\cite{badro_1999} observed signatures of the high-spin state up to 145 GPa at room temperature. Estimates for completion of the transition span a wide pressure range, from 90-140~GPa~\cite{pasternak_FeO} to nearly 200~GPa~\cite{hamada}.

Most investigations of Fe$_{1-x}$O under lower-mantle conditions have been performed using static compression. In contrast, two dynamic compression studies using gas-guns on Fe$_{0.91}$O, Fe$_{0.94}$O, and Fe$_{0.95}$O, reported a volume discontinuity of approximately 4\% at 70-80~GPa~\cite{jeanloz,yagi_1988}, coincident with metallisation near 70~GPa~\cite{knittle_met}. This volume change was interpreted as evidence for a structural phase transition~\cite{jeanloz}, possibly from B1 to B8~\cite{fei_mao,murakami}. However, this assignment remains uncertain, as some alternative determinations of the B8 phase boundary do not intersect the Hugoniot in this pressure range~\cite{fisher,ozawa_2010}, as we show in Fig.~\ref{Fig:phase}.

\begin{figure}
     \centering
     \includegraphics[width=1\columnwidth]{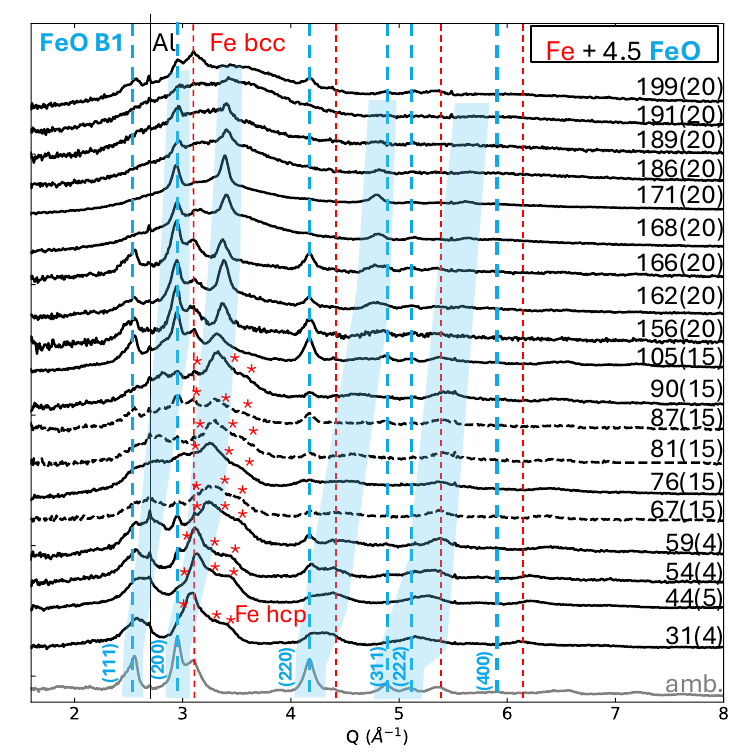}
     \caption{Azimuthally integrated one-dimensional XRD patterns under laser-driven shock compression. Dashed curves correspond to targets in which a sapphire window was glued to the Fe + 4.5 FeO layer. Measurements were obtained in two experiments (p6659 and p6746) with same experimental configuration. Compression of the FeO B1 phase is evidenced by systematic shift of its principal Bragg peaks, highlighted in light blue. Above 191(20)~GPa, the disappearance of peaks is attributed to complete melting of FeO. Iron shows hexagonal close-packed (hcp) structure at 31(4)~GPa; the main hcp reflections (100), (002), and (101) are marked by red stars. Vertical dotted red, dashed cyan, and solid black lines indicate the positions of unshocked bcc-Fe, FeO B1, and Al peaks, respectively.}
     \label{Fig:intensity}
\end{figure}

In this work we retrieve for the first time structural phases along the FeO Hugoniot, and place tighter constrains on the associated density evolution and on the electronic state. We use the Fe-FeO system, which stabilizes nearly stoichiometric FeO under pressure~\cite{stolen,seagle}. Our primary diagnostics are time-resolved XRD and XES, collected in transmission from laser-driven shock-compressed samples. Three experiments (p6746, p6659, and p6656) were performed at the High Energy Density endstation of the European X-ray Free Electron Laser~\cite{zastrau}. For the XRD-only experiments (p6746 and p6659), the x-ray probe operated in self-amplified spontaneous emission (SASE) mode at 24~keV. In the combined XES and XRD experiment (p6656), a seeded x-ray beam probed the sample at 8.3~keV. Details of the set-up are given in Section II of the Supplemental Material (SM). 

The target comprised a 55 or 77~\textmu{}m thick black Kapton (polyimide) ablator and a 10~\textmu{}m thick, homogeneous Fe/Fe$_{1-x}$O layer deposited by Plasma Vapor Deposition (PVD) onto the ablator. ElectronProbe Microbeam Analysis (EPMA) measured a composition of 55.0(0.4)~at\% Fe and 45.0(0.4)~at\% O.
The as-deposited material consisted of intergrown body-centred cubic (bcc) Fe and Fe$_{1-x}$O with the B1 (rocksalt) structure. Assuming FeO stoichiometry, the atomic proportions correspond to Fe + 4.5~FeO. Further discussions on starting material and the impact of possible deviations from FeO stoichiometry are provided in Sections I and IX of the SM. A 200-400 nm Al coating is present on top of the Fe/Fe$_{1-x}$O layer to improve reflectivity.

\begin{figure*}
     \centering
     \includegraphics[width=1\textwidth]{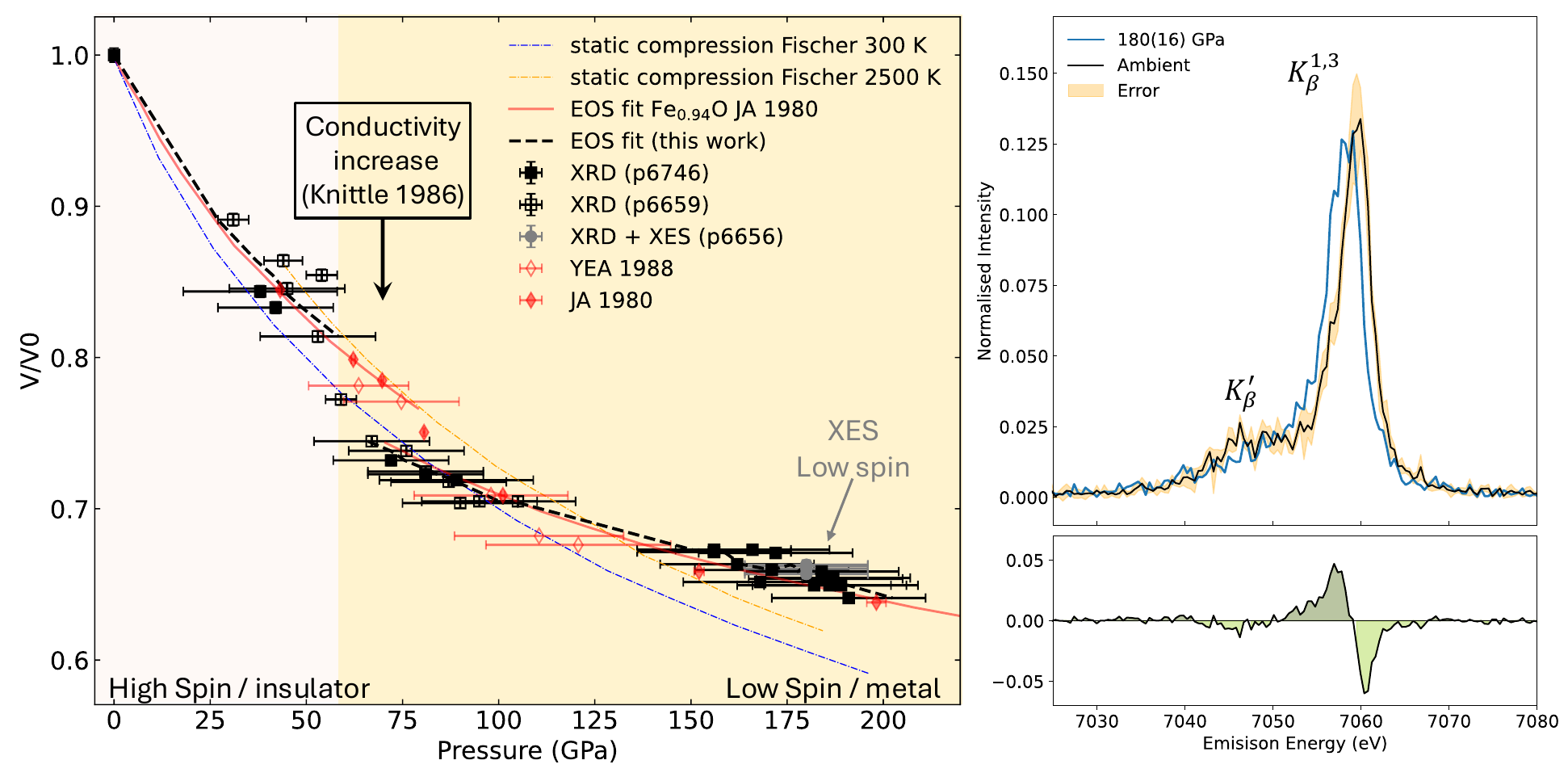}
     \caption{Left: Relative volume $V/V_{0}$ as a function of pressure. The reference molar volume is 11.61~cm$^{3}$.mol$^{-1}$. Previous dynamic compression results include conductivity measurements (Knittle)~\cite{knittle_met} and density data from Jeanloz and Ahrens (JA 1980)~\cite{jeanloz} and Yagi {\it et al.} (YEA 1988)~\cite{yagi_1988}. Equation-of-state fits to static compression data for FeO B1 at 300~K and 2500~K (noted as Fischer for Fischer {\it et al.}~\cite{fisher}) are also plotted. A distinct volume reduction is observed in dynamic compression data between 54(4)-67(15)~GPa in this work, and between 70-80~GPa in JA~\cite{jeanloz}. No comparable discontinuity is evident in static compression measurements within the B1 phase~\cite{fisher,morard2022,fu_hirose,seagle,ozawa_2010,campbell,murakami}. The red solid line shows the equation-of-state (EOS) fit of JA~\cite{jeanloz}, and the black dashed line our EOS fit, constrained by our data below 54~GPa and above 67~GPa. Right: Fe K$_{\beta}$ XES spectra of the Fe + 4.5 FeO sample under ambient conditions and at 180(16)~GPa. The shocked spectrum corresponds to the sum of six shots; the ambient spectrum to the sum of 10 preshots. Error indicates the square root interpolated ambient error. Further details are given in SM, Fig.~S7. Difference spectrum (ambient minus shocked) is shown below. The reduction of the K$_{\beta^{\prime}}$ satellite intensity and the shift of the K$_{\beta_{1,3}}$ main line are consistent with a low-spin (LS) state at 180(16)~GPa.}
     \label{Fig:V_V0}
\end{figure*}

Shock pressure was determined using Velocity Interferometry System for Any Reflector (VISAR)~\cite{barker,descamps_VISAR,hplf}, which provides the shock breakout time and the free-surface velocity history. For shots in which the velocity history could not be recovered, pressures were inferred from an empirical pressure–breakout time relationship (Fig.~S10), calibrated using shots with reliable VISAR data. Complementary VISAR-only shots were taken at the High Power Laser Facility, ID24ED, ESRF synchrotron~\cite{hplf}. At higher pressures, one-dimensional hydrodynamic simulations were performed using the code~\textsc{MULTI}~\cite{ramis}. 

As shown in Fig.~\ref{Fig:intensity}, the FeO-related reflections are well described by the B1 structure up to melting. The five first principal B1 reflections can be tracked from ambient conditions to 191(20)~GPa without the appearance of additional peaks, exhibiting only a systematic shift to higher scattering vector with increasing pressure. The B8 structure does not reproduce the observed Bragg positions (Fig.~S16 of the SM).
No peak splitting is observed between ambient conditions and 31(4)GPa, within the limits of experimental resolution, which would be expected for a B1 to rB1 transition below 30~GPa. Likewise, no change in peak positions or symmetry is detected near 40~GPa that would indicate a transition to a monoclinic phase (Fig.~\ref{Fig:phase}). We note, however, that diffraction peaks are intrinsically broadened under shock compression. The presence of broad hcp-Fe reflections and unshocked material further limits structural resolution. The FeO B1 molar volume was determined from the positions of the five first principal B1 Bragg reflections indicated in Fig.~\ref{Fig:intensity}. The resulting compression curve is shown in Fig.~\ref{Fig:V_V0}.

Bcc Fe present in the initial sample transforms to hcp Fe under compression, as shown in Fig.~\ref{Fig:intensity}. The derived hcp lattice parameters agree with previous measurements (Fig.~S14). Fe reflections are no longer clearly resolved above 70-90~GPa. This loss of diffraction signal may reflect a reduction in hcp-Fe domain size of order 2-15~nm~\cite{hawreliak2008}, as well as strain at Fe–FeO interfaces that degrades crystallinity near grain boundaries. Thus, although diffuse scattering is observed from 166(20)~GPa, we cannot unambiguously attribute it to FeO melt due to possible overlapping Fe signal.

We performed in situ XES measurements under shock compression at 180(16)~GPa and recorded the Fe K$_{\beta}$ emission from the 3p$\rightarrow$1s transition. The spectral shape and the energy splitting between the main K$_{\beta_{1,3}}$ line and the K$_{\beta^{\prime}}$ satellite depend on the exchange interaction between the 3p and 3d electrons, rendering the technique highly sensitive to the Fe spin state, that is, the occupancy of the 3d $e_g$ and $t_{2g}$ levels. A shift of the K$_{\beta_{1,3}}$ line to lower emission energy alongside a reduction in the K$_{\beta^{\prime}}$ satellite intensity is characteristic of a low-spin (LS) state~\cite{peng_1994,vanko,lafuerza}.
Fig.~\ref{Fig:V_V0} shows that the sample is in a high-spin (HS) state at ambient pressure. Under shock compression at 180(16) GPa, the spectrum exhibits a clear reduction of the K$_{\beta^{\prime}}$ satellite and a shift of the K$_{\beta_{1,3}}$ line consistent with a LS state. XES measurements on pure Fe report suppression of the satellite peak above the bcc–hcp transition at 15~GPa, indicative of an LS state in hcp-Fe under pressure~\cite{rueff_Fe}, although an alternative study places the HS–LS boundary above 50~GPa with strong temperature dependence~\cite{ono_Fe}. In the present case, the absence of an HS signature at 180(16)~GPa is consistent with LS Fe; however, the low Fe fraction in the sample precludes a definitive assignment within the statistical accuracy of the experiment.

Our experimental results indicate that no structural phase transition occurs along the Hugoniot: FeO remains in the B1 structure up to melting. This behaviour is consistent with the static compression phase diagram of the Fe–FeO system shown in Fig.~\ref{Fig:phase}. The pressure–volume evolution broadly follows previous static compression measurements for FeO B1~\cite{fisher,seagle,ozawa_2010,campbell,fu_hirose,morard2022}. However, a 7–10\% volume reduction is observed here between 54(4)-67(15)~GPa. A comparable $\sim$4\% volume drop at 70-80~GPa was reported in previous gas-gun experiments~\cite{yagi_1988,jeanloz}, where velocity history was retrieved to extract densities via the Rankine–Hugoniot relations.
The volume discontinuity observed in the present work cannot be attributed to a structural phase transition, as FeO is shown to remain in the B1 phase along the Hugoniot up to melting, contrary to earlier interpretations~\cite{jeanloz,murakami,fei_mao}.

It has also been proposed that the volume jump observed in gas gun 
measurements could reflect a change in Fe$_{1-x}$O 
stoichiometry~\cite{liu_1982}. This would require a decrease from 
Fe$_{0.97}$O to Fe$_{0.75}$O at 220~GPa, as well as a 
pressure/temperature-dependent shift in stoichiometry evolution to 
produce the observed discontinuity~\cite{liu_1982}. However, 
stoichiometry variations above 20~GPa and 1000~K have never been observed to drive the composition below Fe$_{0.9}$O~\cite{mccammon}, and in Fe-excess conditions Fe$_{1-x}$O remains close to stoichiometric~\cite{stolen}. Furthermore, no static compression experiment on Fe~+~Fe$_{1-x}$O as a starting material has reported a major stoichiometry change at conditions comparable to our study~\cite{fisher,morard2022,fu_hirose,seagle,ozawa_2010,campbell,murakami}.

The observed volume discontinuity can therefore be plausibly explained by an electronic transition. Indeed, an insulator-to-metal transition in FeO has previously been reported under both dynamic~\cite{knittle_met} and static compression as shown in Fig.~\ref{Fig:phase}~\cite{fischer_met,ohta}. In addition, a HS$\rightarrow$LS transition has been proposed between 70-200~GPa, although the precise boundaries remain debated. Our XES measurements identify an LS state at 180(16)~GPa under shock compression. We therefore attribute the 7–10\% volume reduction observed between 54(4)-59(4)~GPa and 67(15)~GPa to a spin transition in FeO B1 from HS to LS, potentially accompanied by metallisation. We note that DFT+DMFT calculations predict a volume collapse of up to 9\% associated with the spin transition in FeO B1~\cite{greenberg_FeO,leonov}, in good agreement with our experimental observations. In contrast, smaller volume reductions between 1.6\% and 2.5\% have been reported for spin transitions in the rB1~\cite{ono} and B8 phases~\cite{ozawa_2011}, respectively.

We modeled the compression using a second-order Birch-Murnaghan and Mie-Gr\"uneisen equation of state (Section XII of the SM). Independent fits were performed for pressures up to 54~GPa (59~GPa), corresponding to the HS insulating B1 phase, and pressures above 59~GPa (67~GPa), corresponding to the LS and potentially metallic B1 phase. 
Temperatures were determined by DFT calculations (Section XIII of the SM) along the FeO B1 Hugoniot, displayed in Fig.~\ref{Fig:phase}. The ambient Debye temperature was fixed at 417~K~\cite{stixrude}.
For the HS phase, the reference volume was fixed to the measured ambient value, 11.61(0.03)~cm$^{3}$mol$^{-1}$, and the ambient Gr\"uneisen parameter was set to 1.41~\cite{fisher}. The resulting zero-pressure bulk modulus is 161(12)-177(11)~GPa.
In the high-pressure region, the ambient Gr\"uneisen parameter was fixed at 1.8, consistent with the high-pressure analysis of Jeanloz and Ahrens~\cite{jeanloz}. The fit yields a zero-pressure bulk modulus of 252(20)-266(22)~GPa.

The HS bulk modulus obtained here exceeds the 149~GPa reported from static compression of FeO in the Fe–FeO system~\cite{fisher}, but lies within the broader range of 142-185~GPa determined in earlier studies~\cite{mccammon,jeanloz}. Consistent with DFT+DMFT predictions~\cite{greenberg_FeO,leonov}, the bulk modulus of the LS B1 phase exceeds that of the HS B1 phase.
Although a distinct volume discontinuity is observed here, in agreement with previous dynamic compression studies, no comparable volume drop has been reported under static compression for the B1 phase, neither in the Fe–FeO system nor in non-stoichiometric Fe$_{1-x}$O, despite measurements up to 200~GPa~\cite{fisher,morard2022,fu_hirose,seagle,ozawa_2010,campbell,murakami}. Moreover, reported volume changes associated with spin transitions in the B8 and rB1 phases are at least four times smaller than the 7–10\% discontinuity observed here~\cite{ozawa_2011,ono}.

The occurrence of a volume drop in FeO B1 only under dynamic compression may be due to the timescales involved. Laser-driven shocks probe nanosecond timescales, gas-gun experiments microseconds, whereas static compression experiments probe seconds to hours. The HS state in FeO is stabilised by strong Fe–Fe exchange interactions~\cite{lin_spin}, which also account for the lower transition pressures observed in (Fe,Mg)O (40-80~GPa)~\cite{lin_spin,fei_FeMgO,zhang_FeMgO,speziale,mao_2011,badro_2003}. A time-dependent re-equilibration of the spin state, with progressive reduction of the LS fraction, could therefore suppress an observable volume discontinuity under static conditions. This scenario is consistent with the absence of a volume drop in static experiments, the $\sim$4\% reduction reported in gas-gun measurements~\cite{yagi_1988,jeanloz}, and the larger 7–10\% discontinuity observed here on nanosecond timescales.

In laser-heated diamond anvil-cell experiments, temperature gradients around the hotspot may also influence the measured response. Spatial averaging over regions with different temperatures could mix HS and LS contributions within the probed volume, particularly given the strong temperature dependence of the spin-transition pressure~\cite{greenberg_FeO}.
For (Fe,Mg)O, a density increase of 2–5\% has been reported under both static~\cite{fei_FeMgO,speziale,mao_2011} and dynamic compression~\cite{zhang_FeMgO}. The greater stability of the LS state in the presence of Mg may reduce sensitivity to kinetic effects and temperature, yielding more consistent behaviour across compression and heating pathways. The present results for FeO provide a further example of divergence between static and dynamic compression. Recently, it was shown in Fe$_2$O$_3$ that structural transitions can be kinetically inhibited under shock, while the spin transition remains unaffected~\cite{amourettiPRL}. Here, we demonstrate that the dynamic compression mechanism can also impact electronic transitions and, consequently, can affect the lattice response.

In summary, our results show that FeO remains in the B1 structure along the Hugoniot up to melting, while exhibiting a pronounced volume collapse associated with a spin transition under dynamic compression. The magnitude of this discontinuity, absent in static measurements, highlights the critical role of compression and heating pathway and timescale in governing coupled electronic and structural responses at extreme conditions.

\section{Acknowledgements}

C.Cr\'episson., A.Coutinho Dutra, J.S.Wark, and S.M.Vinko acknowledge support from EPSRC under research grant EP/W010097/1.
M.Fitzgerald, Y.Wang, G.Gregori, and S.M.Vinko acknowledge support from EPSRC and First Light Fusion under the AMPLIFI prosperity partnership, grant EP/X025373/1.
P.G.Heighway and J.S.Wark acknowledge support from EPSRC under research grant EP/X031624/1.
D.J.Peake, H.Taylor, and T.Stevens acknowledge support from AWE via the Oxford Centre for High Energy Density Science (OxCHEDS).
K.Appel, C. Sternemann, and C.Camarda acknowledge the financial support by Deutsche Forschungsgemeinschaft (DFG, German Research Foundation) via project AP262/3-1 and STE1079/10-1 (project number 521549147) and financial support from the Federal Ministry of Research, Technology and Space (BMFTR) under project ID 05K19PE2. A.Amouretti, K.Yamamoto, and N.Ozaki acknowledge support from Japan Society for the Promotion of Science (JSPS) KAKENHI (Grant Nos. 23K20038 and 25H00618), JSPS Core-to-Core Program (JPJSCCA20230003), and MEXT Quantum Leap Flagship Program (JPMXS0118067246). E.E.McBride. and A.Descamps were supported by the UK Research and Innovation Future Leaders Fellowship (Grant No. MR/W008211/1) awarded to E.E.McBride. 
N.J.Hartley was supported by the DOE Office of Science Fusion Energy Science under FWP No. 100182.
B.Nagler is supported by U.S. Department of Energy Office of Science, Fusion Energy Science, under Contract No. DE-AC02-76SF00515.
J.D.McHardy, C.V.Storm and M.I.McMahon acknowledge support from EPSRC under research grant Nos. EP/R02927X/1 and EP/Z533671/1. C.M.Lonsdale is grateful to S.G. MacLeod and AWE for the award of their studentship WT13356629. 
J.L\"utgert was supported by GSI Helmholtzzentrum für Schwerionenforschung, Darmstadt, as part of the R\&D project SI-URDK2224 with the University of Rostock.
C.A.J.Palmer and C.Prestwood acknowledge the support of STFC XFEL Physical Sciences Hub.
A.Higginbotham acknowledges support from EPSRC under grants  EP/S023585/1 and EP/X025373/1.
Contributions from L.Dresselhaus-Marais, A.Hari and S.Parsons were supported by the Department of Energy, National Nuclear Security Administration Center of Excellence CAMCSE under Award No. DE-NA0004154.
U.Trdan acknowledges the financial support from the Slovenian Research Agency: Research core funding No. P2-0270 and projects No. J2-60033 (SuperShocked) and and N2-0328 (Weave-NCN). U.Trdan would also like to thank the Horizon Europe framework’s Sustainable Blue Economy Partnership for funding the CORRASBlue project. 
A.Chakraborti is supported by the Agence Nationale de Recherche (ANR, France) projet MIN-DIXI.
S.Merkel, J.Chantel, and H.Ginestet are funded by the European Union (ERC, HotCores, Grant No. 101054994).
This work was supported by Deutsche Forschungsgemeinschaft (DFG) Project Nos. 495324226 (S.Schumacher and D.Kraus) and 505630685 (J.Kuhlke and D.Kraus). This research was supported by the Centre for Molecular Water Science (CMWS) in an Early Science Project (D.Ranjan). D.Kraus is supported by the European Union (ERC, MEGACHEM, Grant No. 101171289). 
Views and opinions expressed are however those of the author(s) only and do not necessarily reflect those of the European Union or the European Research Council. Neither the European Union nor the granting authority can be held responsible for them.
J.D.Umpleby-Thorp. and A-M.Norton acknowledge support from EPSRC under grant EP/S022430/1, Centre for Doctoral Training in Fusion Energy Science and Technology.
S.Pandolfi acknowledges support from the ANR grant HEX-DYN (ANR-24-CE30-4792).
This result is part of a project that has received funding from the European Research Council (ERC) under the European Union’s Horizon 2020 research and innovation programme (Grant agreement No. 101002868).
Part of this work was performed under the auspices of the US Department of Energy at Lawrence Livermore National Laboratory
under Contract DE-AC52-07NA27344 (LLNL-JRNL-819849). Part of this work was funded by LDRD 24ERD-021.
The work at University of South Florida from I.I.Oleynik, F.Hanby, A.Tipeev, C.N.Somarathna, J.D.Tunacao, and S.Galitskiy is supported by DOE/NNSA (Award No. DE-NA0004234), DOE/FES (Award No. DE-SC0024640), NSF (Award No. 2421937), and NASA (Award No. 80NSSC25K7172). 
This work was partially supported by the National Natural Science Foundation of China (Grant No. 12402466, 11627901).   
This work was partially supported by the Center for Advanced Systems Understanding (CASUS), financed by Germany’s Federal Ministry of Education and Research (BMBF) and the Saxon state government out of the State budget approved by the Saxon State Parliament.
This work has received funding from the European Union's Just Transition Fund (JTF) within the project \textit{R\"ontgenlaser-Optimierung der Laserfusion} (ROLF), contract number 5086999001, co-financed by the Saxon state government out of the State budget approved by the Saxon State Parliament.
This material is based upon work supported by the Department of Energy [National Nuclear Security Administration] University of Rochester “National Inertial Confinement Fusion Program” under Award Number(s) DE-NA0004144. 

Part of Al coating was realized at DESY, Hamburg, Germany by Michael R\"{o}per. Samples were measured using the Co x-ray diffractometer and Electron MicroProbe at the Department of Earth Sciences at the University of Oxford assisted by K.Sokol and A.Matzen.
Sample preparation for analysis was performed at the polishing lab at the Department of Earth Sciences at the University of Oxford assisted by E.Donald. 
We acknowledge the European Synchrotron Radiation Facility (ESRF) for provision of synchrotron radiation facilities under proposal numbers ES1629 and ES1532. Some of the experimental results were obtained during proposal number 6746, during collaborative proposal 6656 (Main proposer: Dominik Kraus; Principal investigator: Thomas Preston), and during collaborative proposal 6659, carried out in November 2024 at EuXFEL (Main Proposer: Guillaume Morard; Principal Investigator: Jon Eggert; Local Contact: Karen Appel), with the assistance of pillar leaders Adrien Descamps, Andrew Higginbotham, Trevor Hutchinson, Chris McGuire, Silvia Pandolfi and Arnaud Sollier. We acknowledge the European XFEL in Schenefeld, Germany, for provision of X-ray free electron laser beam time at the Scientific Instrument HED (High Energy Density Science) and would like to thank the staff of European XFEL and DESY for their assistance. The authors are indebted to the HIBEF user consortium for the provision of Instrumentation and staff that enabled this experiment.
For the purpose of open access, the author has applied a Creative Commons Attribution (CC BY) license to any Author Accepted Manuscript version arising from this submission.

\section{Data availability}

The data from EuXFEL presented here are available upon reasonable request at https://doi.org/10.22003/XFEL.EU-DATA-006746-00, doi:10.22003/XFEL.EU-DATA-006656-00 once the data embargo of the experiment campaign 6656 has been lifted (20 Aug 2027). Before the end of the data embargo, all relevant data are available from the authors upon reasonable request, and doi:10.22003/XFEL.EU-DATA-006659-00.
Corresponding run numbers are listed in Table 2 of the Supplemental Material.

\end{document}